\documentclass[journal=jacsat,manuscript=article]{achemso}
\pdfoutput=1

\usepackage[version=3]{mhchem} %
\usepackage[utf8]{inputenc} %
\usepackage[T1]{fontenc} %
\usepackage{graphicx}       %
\usepackage{amsmath, amssymb}        %
\usepackage{breakcites}     %
\usepackage{array}          %
\usepackage{multirow}       %
\usepackage{threeparttable}
\usepackage{pdflscape}
\usepackage{xcolor}
\usepackage{bibunits}
\SectionNumbersOn

\graphicspath{ {./images/} }

\newcommand{\red}[1]{\textcolor{red}{#1}}
\author{Thijs Stuyver}
\affiliation[MIT]
{Department of Chemical Engineering, Massachusetts Institute of Technology, 77 Massachusetts Avenue, Cambridge, Massachusetts 02139, United States}
\author{Connor W. Coley}
\affiliation[MIT]
{Department of Chemical Engineering, Massachusetts Institute of Technology, 77 Massachusetts Avenue, Cambridge, Massachusetts 02139, United States}
\altaffiliation{Department of Electrical Engineering and Computer Science, Massachusetts Institute of Technology, 77 Massachusetts Avenue, Cambridge, Massachusetts 02139, United States}
\email{ccoley@mit.edu}

\title{Quantum chemistry-augmented neural networks for reactivity prediction: Performance, generalizability and interpretability}

\abbreviations{}

\begin{document}

\begin{tocentry}

\includegraphics[scale=0.97]{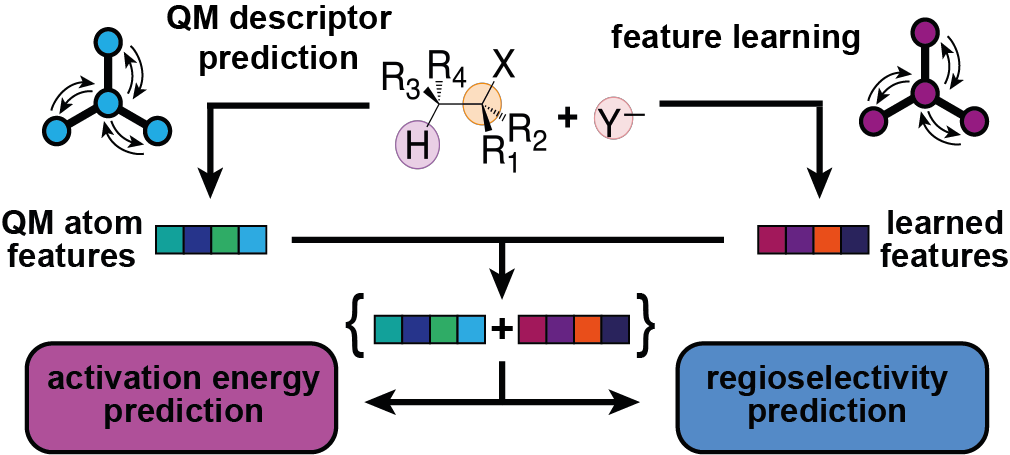}

\end{tocentry}

\begin{abstract}
There is a perceived dichotomy between structure-based and descriptor-based molecular representations used for predictive chemistry tasks. Here, we study the performance, generalizability, and interpretability of the recently proposed quantum mechanics-augmented graph neural network (ml-QM-GNN) architecture as applied to the prediction of regioselectivity (classification) and of activation energies (regression). In our hybrid QM-augmented model architecture, structure-based representations are first used to predict a set of atom- and bond-level reactivity descriptors derived from density functional theory (DFT) calculations. These estimated reactivity descriptors are combined with the original structure-based representation to make the final reactivity prediction. We demonstrate that our model architecture leads to significant improvements over structure-based GNNs in not only overall accuracy, but also in generalization to unseen compounds. Even when provided training sets of only a couple hundred labeled data points, the ml-QM-GNN outperforms other state-of-the-art model architectures that have been applied to these tasks. Further, because the predictions of our model are grounded in (but not restricted to) QM descriptors, we are able to relate  predictions to the conceptual frameworks commonly used to gain qualitative insights into reactivity phenomena. This effort results in a productive synergy between theory and data science, wherein our QM-augmented models provide a data-driven confirmation of previous qualitative analyses, and these analyses in their turn facilitate insights into the decision-making process occurring within ml-QM-GNNs.

\end{abstract}

\begin{bibunit}[unsrt]
\section*{Introduction}
The rationalization and prediction of reactivity trends is one of the core objectives of (theoretical) chemistry. Before advanced \emph{ab initio} quantum mechanics (QM) computations became commonplace, a plethora of heuristic concepts and qualitative theory-inspired rules---usually tailored to a specific subclass of compounds or reactions---were already proposed and developed to this end. Some iconic examples in this regard are the Woodward-Hoffmann rules for the prediction of reaction outcomes of pericyclic transformations, \cite{woodward1965stereochemistry} the Bell-Evans-Polanyi principle, \cite{evans1936further} and the hard-soft acid-base (HSAB) concept pioneered by Pearson.\cite{pearson1963hard} 

Later on, several competing, overarching theoretical frameworks emerged, all of which aspire to describe chemical reactions in a universally applicable manner. Some well-known examples of such frameworks are conceptual density functional theory (cDFT), \cite{parr1983absolute,geerlings2003conceptual} the (molecular orbital-based) activation-strain model (ASM) \cite{fernandez2014activation,bickelhaupt2017analyzing}, and the valence bond (VB) reactivity model. \cite{shaik1981happens, shaik1999valence} At the core of each of these frameworks is the definition of a limited set of chemically meaningful quantities or descriptors. These probe,  either directly or indirectly, the magnitude of different fundamental interactions  (e.g., orbital-based or "soft-soft" vs. electrostatic or "hard-hard" interactions in cDFT), which collectively characterize the chemical system and enable an insightful and internally consistent (qualitative) discussion of its reactivity. As these frameworks matured, a concerted effort has taken place to incorporate/embed the various previously proposed concepts and rules, albeit of limited scope, into them. \cite{parr1984density,shaik2007chemist} Attempts have also been made to build bridges between the individual frameworks themselves, thus facilitating enhanced understanding. \cite{hoffmann2003conversation, borden2017dioxygen, stuyver2020local, stuyver2020unifying}

With the advent of machine learning (ML) and artificial intelligence methods, an entirely different approach to chemical reactivity prediction  emerged, not constrained by any specific theoretical framework. Recent work in this area includes the prediction of reaction products, \cite{wei2016neural,coley2017prediction,nair2019data} reaction yields, \cite{zuranski2021predicting,ahneman2018predicting,schwaller2021prediction} and bond dissociation energies. \cite{john2020prediction,wen2021bondnet} Instead of building physically motivated representations of the constituent molecules (e.g., of reactants, products, catalysts), many ML approaches take advantage of a model's ability to learn meaningful representations and start with simple structural descriptors/features. These include graph-based molecular representations using graph neural networks (GNNs) \cite{duvenaud2015convolutional,coley2019graph,john2020prediction,wen2021bondnet}, simpler molecular fingerprint\cite{rogers2010extended} representations,\cite{wei2016neural, sandfort2020structure} and even SMILES representations. \cite{schwaller2019molecular,schwaller2021prediction,yang2019molecular} Such structure-based representations do not necessarily have a \textit{direct} connection to reactivity and rely on the nonlinearity and expressivity of ML models to relate structure to function. \cite{Goodfellow-et-al-2016} Nevertheless, given sufficient data to train these models, deep learning for reactivity prediction has been demonstrated to achieve accurate predictions of reaction products and/or energies.

Unfortunately, the setup of regular GNNs renders the decision-making process occurring inside them rather opaque. Rationalizing the predictions made by these networks has largely been limited to brittle techniques for estimating the sensitivity of predictions to atom- and bond-level contributions. \cite{polishchuk2017interpretation, ahneman2018predicting, niemeyer2016parameterization, amar2019machine, wu2017parameterization}  Therefore, these models have often been characterized as black boxes. Furthermore, structure-based GNNs tend to/are assumed to underperform in data-limited settings as they must learn a meaningful representation ``from scratch''. \cite{DBLP:journals/corr/abs-2011-12203, friederich2020machine} This represents a significant drawback of GNNs, as large data sets with thousands of data points are rare in the field of chemistry, especially when focusing on subtle reactivity questions. \cite{gallegos2021importance} Other machine learning methods, e.g., kernel ridge regression (KRR) \cite{meyer2018machine,heinen2020quantum} or Gaussian process (GP) regression\cite{friederich2020machine} in combination with  molecular representations such as bag-of-bonds (BoB), \cite{hansen2015machine} FCHL19 \cite{christensen2020fchl} or spectrum of London and Axilrod-Teller-Muto potentials (SLATM) \cite{axilrod1943interaction}  are generally assumed to perform better in the face of data scarcity; even then, several thousand data points may be  needed to train a model to an acceptable level of accuracy. \cite{meyer2018machine, friederich2020machine,heinen2020quantum}

One strategy to mitigate the drawbacks of these data-hungry methods is to represent molecules with functional descriptors that have a more direct (and linear) relationship with their reactivity. Instead of working within a specific theoretical framework,  reactivity is modelled through statistical methods, but descriptors/features used as input are inspired by the chemistry and physics underlying the investigated reactivity problem. Champions of this approach are among others Sigman and co-workers, who have employed multivariate linear regression to relate sophisticated electronic and steric descriptors to complex properties such as enantioselectivity, \cite{sigman2016development} and Denmark et al., who used support vector machines (SVM) and feed-forward neural networks for the same task. \cite{zahrt2019prediction} Recently, Doyle and co-workers demonstrated the prediction of reaction yields of C-N cross-coupling reactions with a random forest model by selecting reaction-specific descriptors, starting from approximately 4,000 data points obtained via high-throughput experimentation. \cite{ahneman2018predicting} Other recent examples of this strategy can be found in the work of Beker et al.,\cite{beker2019prediction} Hong and co-workers \cite{li2020predicting} and Jorner et al. \cite{jorner2021machine} 

While descriptor-based methods may require less data, be more generalizable, i.e., achieve an improved performance on classes of compounds not present in the training set, \cite{beker2019prediction} and enable at least some interpretability compared to the use of universal/non-specific representations, \cite{estrada2018response} the selection of suitable descriptors is a non-trivial and problem-specific matter. \cite{gallegos2021importance} Even more importantly, in order to obtain these chemically-relevant descriptors, a data set specific computational workflow is often required, which creates a bottleneck that significantly hampers the ease of employability of this strategy. \cite{gallegos2021importance}

In this work, we build upon a unifying approach proposed by Guan et al.\cite{guan2021regio}, which aims to combine the advantages of structure-based GNNs with those of models based on expert-guided descriptors. Instead of feeding computationally expensive, task-specific  descriptors directly to a machine learning model, we start from graph-based/structural input features but predict, as an intermediate step, a set of atom- and bond-level QM descriptors prior to the final reactivity prediction. Taking this approach, it becomes possible to construct a QM-based representation on-the-fly at minimal cost. 

We assess model performance on computational and experimental data sets (focusing on competing E2 vs. S$_\text{N}2$ reactions \cite{von2020thousands} and aromatic substitution reactions \cite{guan2021regio}, respectively) and find that the resulting model architecture exhibits excellent accuracy in a data-limited regime. We further evaluate the model's ability to generalize to structures not seen during training using non-random data splits and observe a comparable improvement in performance.  Importantly, since our ml-QM-GNN models base their final reactivity predictions on a representation partially comprising QM descriptors, we are able to build a bridge to the traditional theoretical reactivity frameworks and interpret the neural network's decision-making process in terms of traditional chemistry concepts. Overall, our work underscores that machine learning techniques and conceptual models are not mutually exclusive approaches, but are able to benefit each another in a synergistic manner.

\section*{Computational Methods}

\subsection*{Model architecture}
\begin{figure}[H]
\centering
\includegraphics[scale=0.7]{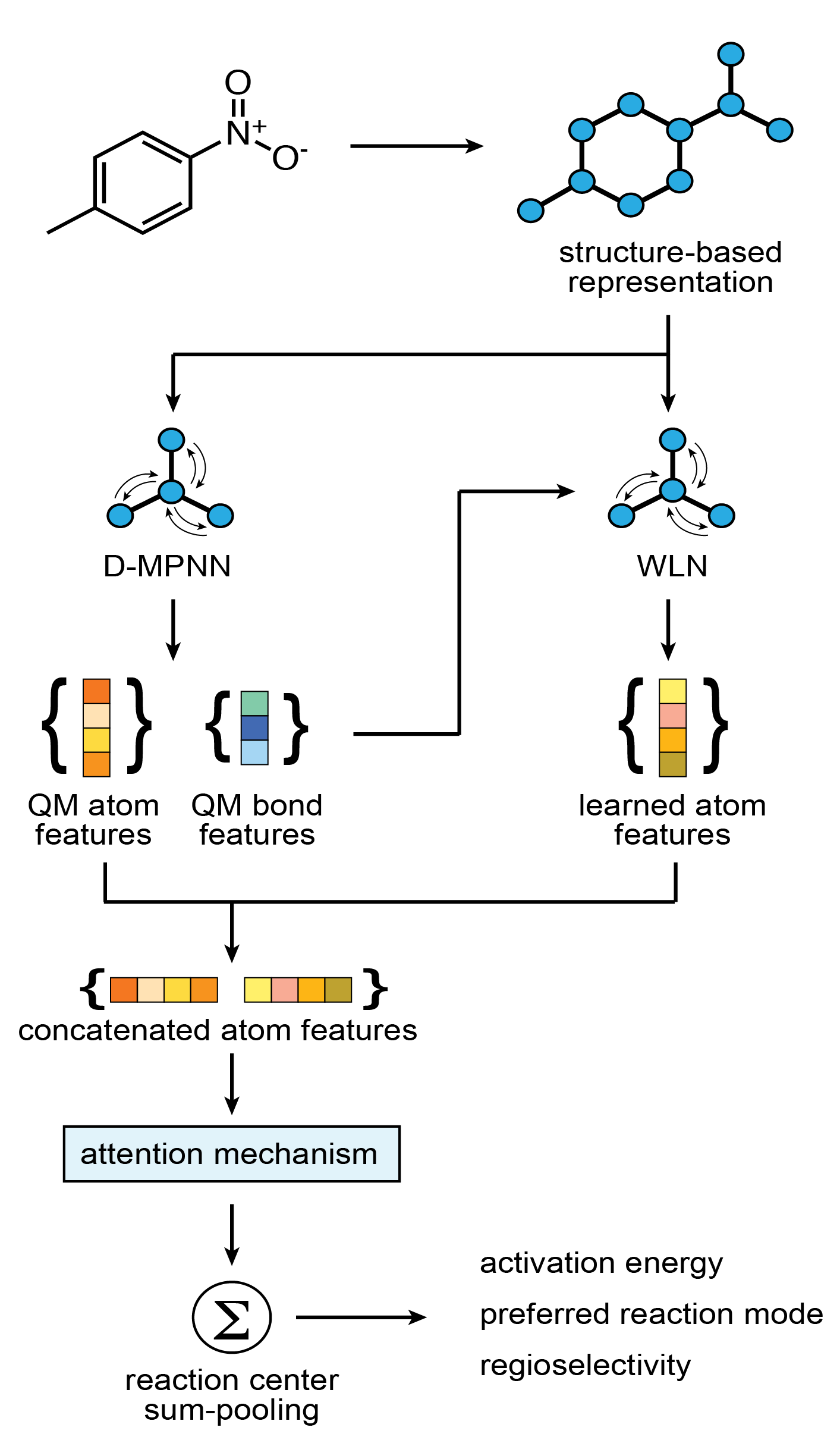}
\caption{A schematic overview of the ml-QM-GNN model architecture. WLN denotes the Weisfeiler-Lehman network branch and D-MPNN denotes the directed message-passing branch of the network.}
\label{scheme_ml_QM_GNN}
\end{figure}

A schematic overview of the complete QM-augmented neural network architecture used in this work is presented in Figure \ref{scheme_ml_QM_GNN} (cf. Section \ref{sec:architecture} in the Supporting Information for an in-depth discussion of the individual network branches).

First, the simplified molecular-input line-entry system (SMILES) representations of compounds involved in the reaction are parsed into graph-based representations using RDKit, \cite{landrum2006rdkit} and structural descriptors (atomic number, formal charge, ring status, bond order, etc.) are calculated for each heavy atom and bond. The structure-based representation is then used in first instance as input for a multitask GNN for QM descriptor prediction, based on a directed message passing neural network (D-MPNN) encoder, which has been adopted without modification from Guan et al. \cite{guan2021regio} (details related to the setup  and training of this D-MPNN encoder are included in Section \ref{sec:architecture}). 

The set of descriptors predicted by the D-MPNN encoder can be subdivided in two main categories: bond descriptors, i.e., natural population analysis (NPA) bond orders \cite{glendening2013nbo} and bond lengths, and atom-centered descriptors, i.e., (Hirshfeld) atomic charges, \cite{hirshfeld1977bonded} as well as (atom-condensed) nucleo- and electrophilic Fukui functions, \cite{geerlings2003conceptual} and NMR shielding constants. \cite{wolinski1990efficient} In the finite-difference approximation, \cite{yang1986use} the electrophilic Fukui function on site i ($f^+_i$) is defined as
\begin{equation}
f^+_i = q_i(N) - q_i(N+1)
\end{equation}
where $q_i(N)$ and $q_i(N+1)$ are the (Hirshfeld) partial charges on atom $i$ for the corresponding (N)- and (N+1)-electron wave function evaluated in the optimized N-electron geometry. 
The nucleophilic Fukui function ($f^-_i$) in its turn is defined as
\begin{equation}
f^-_i = q_i(N-1) - q_i(N)
\end{equation}
where $q_i(N)$ and $q_i(N-1)$ are the (Hirshfeld) partial charges on atom $i$ for the corresponding (N)- and (N-1)-electron wave function evaluated in the optimized N-electron geometry.

The bond descriptors are converted to a vector representation through the application of a radial basis function (RBF) expansion and relayed to a separate Weisfeiler-Lehman network (WLN) branch, which combines them with the initial, structure-based representation in a convolutional embedding. \cite{NIPS2017_ced556cd} The atom-centered descriptors are kept separately and similarly post-processed, i.e., they are scaled and then turned into a vector representation through RBF expansion.

The learned representation emerging from the WLN and the expanded atomic QM descriptor representation emerging from the D-MPNN branch of the network are subsequently concatenated, after which the concatenated representation is passed through a dense activation layer followed by a global attention mechanism \cite{yang2016hierarchical} to capture the influence of distant parts of the reacting system. Because we focus on molecule- or reaction-level prediction tasks in this work, we aggregate these representations into one final feature vector by sum-pooling  over the (hypothetically) reacting atoms. This global feature vectors is transformed once more in a single-layer network to produce the barrier heights/activation energies in the case of a regression task and a regioselective preference in the case of a classification task (\emph{vide infra}).

The regular GNN, used as a baseline in the discussions below, follows a similar architecture as the ml-QM-GNN, except that it does not contain the QM descriptor network branch; i.e., the WLN operates on the original structural features alone (cf. Section \ref{sec:architecture} in the Supporting Information).
 
\subsection*{Data sets}

\begin{figure}[H]
\centering
\includegraphics[scale=0.6]{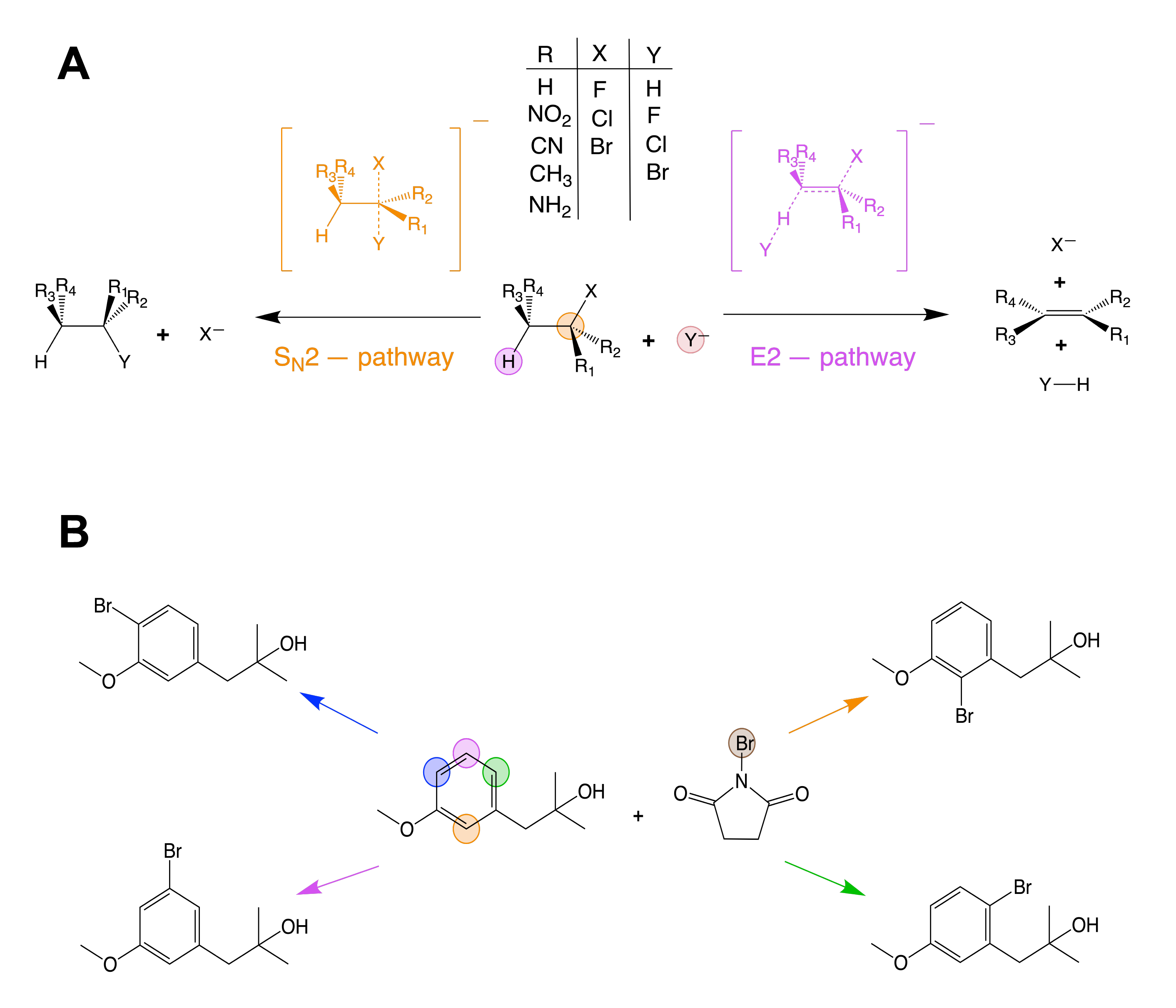}
\caption{Schematic representation of the two considered data sets. (a) The competing E2/S$_\text{N}2$ reaction pathways, with the respective attacking positions of the nucleophile indicated in purple/orange respectively. The top center table gives an overview of the different substituents (R), leaving groups (X) and nucleophiles (Y) present in the data set.  (b) An example of a data point in the aromatic substitution reaction data set. Colored dots indicate potential reacting sites.}
\label{fig:datasets}
\end{figure}

We focus our evaluation of the ml-QM-GNN on two publicly-available data sets: one computational and one experimental. The first comprises computed stationary points along the potential energy surface for competing E2 and S$_\text{N}2$ reactions in the gas-phase, recently published by Von Lilienfeld and co-workers. \cite{von2020thousands} In total, four distinct nucleophiles (H$^-$, F$^-$, Cl$^-$ and Br$^-$), three distinct leaving groups (F$^-$, Cl$^-$, Br$^-$) and permutations of five potential substituents (H, NO$_2$, CN, CH$_3$ and NH$_2$), on an ethyl-based scaffold, were considered (Figure \ref{fig:datasets}a). 

To construct a regression task, 3647 barrier heights calculated at DF-LCCSD/cc-pVTZ// MP2/6-311G(d) level-of-theory \cite{kendall1992electron,hampel1992comparison,frisch1984self,schutz2003linear,mclean1980contracted,krishnan1980self,dunning1989gaussian} were extracted (1286 corresponding to E2 pathways; 2361 to S$_\text{N}2$). \cite{heinen2020quantum} An appropriate input for our GNNs was constructed by converting the 3D reactant complex geometry for each reaction into a 2D SMILES representation using xyz2mol.\cite{jensen2020xyz2mol} Additional details about the data pre-processing can be found in Section \ref{sec:duplicate}. 

To construct a classification task, we extracted  both an E2 and S$_\text{N}2$ transition state (TS) from 791 unique reaction systems (Section \ref{sec:classification}). These data were used for training classification models to predict whether the E2 or S$_\text{N}2$ pathway is kinetically favored.

The second data set we examine is a purely experimental one consisting of regioselective aromatic C-X substitution and C-H functionalization reactions from the Pistachio database, to which plausible side products corresponding to regiochemical alternatives, identified through template extraction and application \cite{coley2019rdchiral}, were added. \cite{Pistachio} This data set was originally curated by Guan et al. \cite{guan2021regio} Since the data in Pistachio is only available to license-holders, Guan and co-workers filtered out a subset of 3242 data points for which the reactions are also present in the USPTO public database. Here, we examine  the multi-way classification performance when predicting the regiochemical preference for this subset of reactions (Figure \ref{fig:datasets}b). Models applied to this filtered data set were trained on a sample of only 200 data points (unless explicitly stated otherwise).
 
\section*{Results and discussion}

\subsection*{Accuracy -- E2/S$_\text{N}$2 data set}

As a first step, the performance of our ml-QM-GNN model was assessed by comparing the accuracy obtained for the barrier height/activation energy prediction for the competing E2/S$_\text{N}2$ reactions with the (structure-based) GNN baseline model. Initially, we split up the data according to the respective reaction type and performed three 5-fold cross-validations (CV) on each. In every iteration, the number of labelled data points considered for the construction of the model, i.e., the combination of training and validation set, was limited to 125, 250, 500 and 1000 in the case of E2, and 225, 450, 900 and 1800 points in the case of S$_\text{N}2$, matching the prior evaluation by Heinen et al. \cite{heinen2020quantum} The average mean absolute error (MAE) for our GNN models, obtained in this manner, are presented in the first two columns of Tables~\ref{tbl:SN2-performance} and \ref{tbl:E2-performance}. 

\begin{table}
\centering
\begin{threeparttable}
  \caption{Average MAE (kcal/mol) when predicting S$_\text{N}$2 barrier heights, obtained after three 5-fold CVs, for the baseline GNN and our ml-QM-GNN for different numbers of labeled data points. Standard deviations were determined based on the MAEs for the three replicates. The corresponding accuracies obtained from 5-fold CV for the KRR models combined with BoB, SLATM, FCHL19 and one-hot encoding representations are included as well.\cite{heinen2020quantum} 20\% of labeled points were reserved as a validation set for early stopping while training the GNN models.}
  \label{tbl:SN2-performance}
  \begin{tabular}{cccccccc}
    \hline
    labeled points & \begin{tabular}{@{}c@{}} baseline \\ GNN \end{tabular} & ml-QM-GNN & BoB$^\dagger$ & SLATM$^\dagger$ & FCHL19$^\dagger$ & \begin{tabular}{@{}c@{}}one-hot \\ encoding$^\dagger$ \end{tabular}  \\
    \hline
    225 (180 + 45) & $9.07 \pm 0.04$ & $3.61 \pm 0.14$ & 4.89 & 4.44 & 3.80 & 3.53  \\
    450 (360 + 90) & $8.89 \pm 0.13$ & $3.28 \pm 0.04$ & 4.28 & 3.87 & 3.43 & 2.80 \\
    900 (720 + 180) & $8.61 \pm 0.07$ & $2.97 \pm 0.03$  & 3.78 & 3.21 & 3.11 & 2.42 \\
    1800 (1440 + 360) & $8.49 \pm 0.03$ & $2.76 \pm 0.01$  & 3.49 & 2.92 & 2.87 & 2.14 \\
    \hline
  \end{tabular}
\begin{tablenotes}
\item[$\dagger$] Taken directly from Heinen et al. \cite{heinen2020quantum}
\end{tablenotes}
\end{threeparttable}
\end{table}

\begin{table}
\centering
\begin{threeparttable}
  \caption{Comparison of the average MAE (kcal/mol) values on the predicted E2 barrier heights, obtained after three 5-fold CVs, for the baseline GNN and our ml-QM-GNN for different numbers of labeled data points. Standard deviations were determined based on the MAEs for the three replicates. The corresponding accuracies obtained from 5-fold CV for the KRR models combined with BoB, SLATM, FCHL19 and one-hot encoding representations are included as well. 20\% of labeled points were reserved as a validation set for early stopping while training the GNN models.}
  \label{tbl:E2-performance}
  \begin{tabular}{cccccccc}
    \hline
    labeled points & \begin{tabular}{@{}c@{}} baseline \\ GNN \end{tabular} & ml-QM-GNN & BoB$^\dagger$ & SLATM$^\dagger$ & FCHL19$^\dagger$ & \begin{tabular}{@{}c@{}}one-hot \\ encoding$^\dagger$ \end{tabular}  \\
    \hline
    125 (100 + 25) & $9.03 \pm 0.18$ & $4.08 \pm 0.06$ & 4.67 & 4.43 & 4.01 & 3.53  \\
    250 (200 + 50) & $8.78 \pm 0.35$ & $3.24 \pm 0.08$ & 4.07 & 3.87 & 3.42 & 3.12 \\
    500 (400 + 100) & $8.18 \pm 0.04$ & $2.91 \pm 0.02$ & 3.71 & 3.21 & 3.01 & 2.69 \\
    1000 (800 + 200) & $8.04 \pm 0.17$ & $2.65 \pm 0.02$ & 3.27 & 2.92 & 2.75 & 2.40 \\
    \hline
  \end{tabular}
\begin{tablenotes}
\item[$\dagger$] Taken directly from Heinen et al. \cite{heinen2020quantum}
\end{tablenotes}
\end{threeparttable}
\end{table}

Comparison between the MAEs obtained for the QM-augmented model and for the baseline model reveals that inclusion of the QM descriptors in the model architecture results in a 5-6 kcal/mol lower error. The rate at which the accuracy improves as more labelled data points are included during training/validation is also improved. 

To contextualize the performance of our models, we also include kernel ridge regression models benchmarked by Heinen et al. for 4 different (global) representations -- respectively BoB, SLATM, FCHL19 and one-hot encoding (in which every substituent site, as well as the nucleophile and leaving group site, are assigned a bit vector spanning all the possible substituent/nucleophile/leaving group species) -- in Tables~\ref{tbl:SN2-performance} and \ref{tbl:E2-performance}.\cite{heinen2020quantum}

From these tables, it is straightforward to discern that the QM-augmented model outperforms BoB, SLATM and FCHL19 in combination with KRR for even the smallest number of labeled data points (125 or 225 points). In the absence of the QM augmentation, however, the baseline GNN is significantly worse than any of these methods. Further underscoring the excellent performance of the ml-QM-GNN model is the lack of hyperparameter optimization and the use of the mean square error loss during training, rather than MAE directly (cf. Section \ref{sec:architecture}).  Nevertheless, it should also be noted that simple one-hot encoding combined with KRR still outperforms the QM-augmented model here. However, as will be demonstrated below, one-hot encoding suffers from an inherent limitation related to generalization, which limits its appeal with respect to our ml-QM-GNN. 

\begin{figure}[h]
\centering
\includegraphics[scale=0.6]{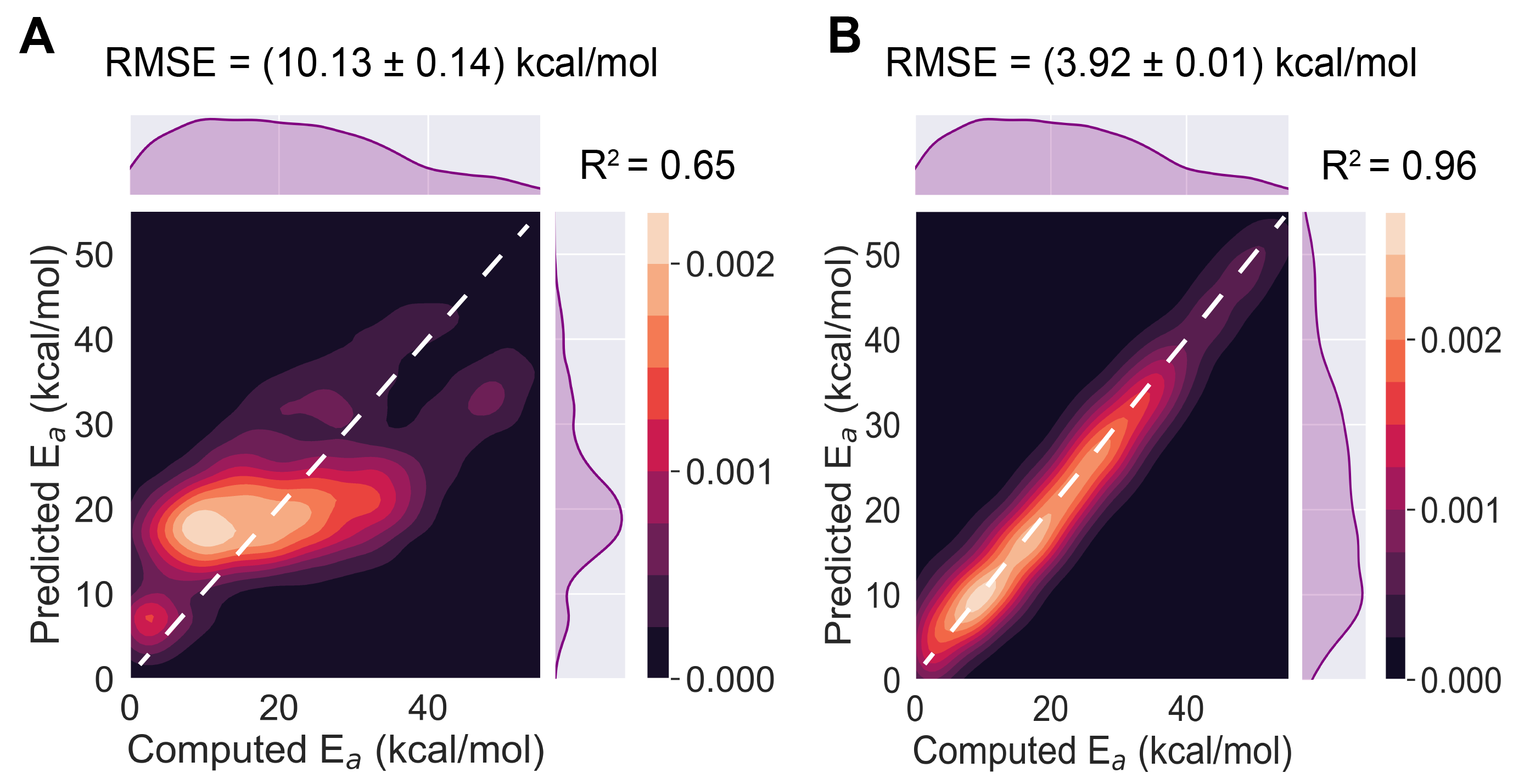}
\caption{Correlation plots for (a) the regular GNN model and (b) the ml-QM-GNN model applied to the full E2/S$_\text{N}$2 activation energy data-set from three 5-fold cross-validations in a 60/20/20-split. The colorbars indicate the scale of the 2D kernel density estimate plots. Standard deviations were determined based on the MAEs for the three replicates.}
\label{Fig_ml_QM_GNN_vs_GNN}
\end{figure}

Since our GNN models involve pooling over reacting atoms, they enable simultaneous treatment of distinct reaction modes for the same reactant/reagent system (in contrast to the reaction-specific KRR models). Hence, we were able to combine the data for the E2 and S$_\text{N}$2 reactions and train a common model for both sets of barrier heights. In Figure \ref{Fig_ml_QM_GNN_vs_GNN}, correlation plots between the "true" (computed) and predicted activation energies (E$_a$), aggregated across all test sets sampled during three 5-fold CVs, are presented.

The correlation between the predicted and computed values is weak for the baseline GNN model (Figure \ref{Fig_ml_QM_GNN_vs_GNN}a; $R^2 = 0.65$); the mean MAE and RMSE from three replicates is 8.4 and 10.1 kcal/mol respectively. The QM-augmented model achieves much stronger correlation (Figure \ref{Fig_ml_QM_GNN_vs_GNN}b; $R^2 = 0.96$), and an average MAE and RMSE of 2.9 and 3.9 kcal/mol.

\subsection*{Generalizability -- E2/S$_\text{N}$2 regression data set}

Next, we considered the ability of our regression models to generalize to new compounds not present in the training data. While one-hot encoding representations have proven to be well suited for reactivity problems focused on interpolation, \cite{granda2018controlling,chuang2018comment} they do not perform well on in out-of-sample predictions. \cite{estrada2018response}

\begin{figure}[h]
\centering
\includegraphics[scale=0.58]{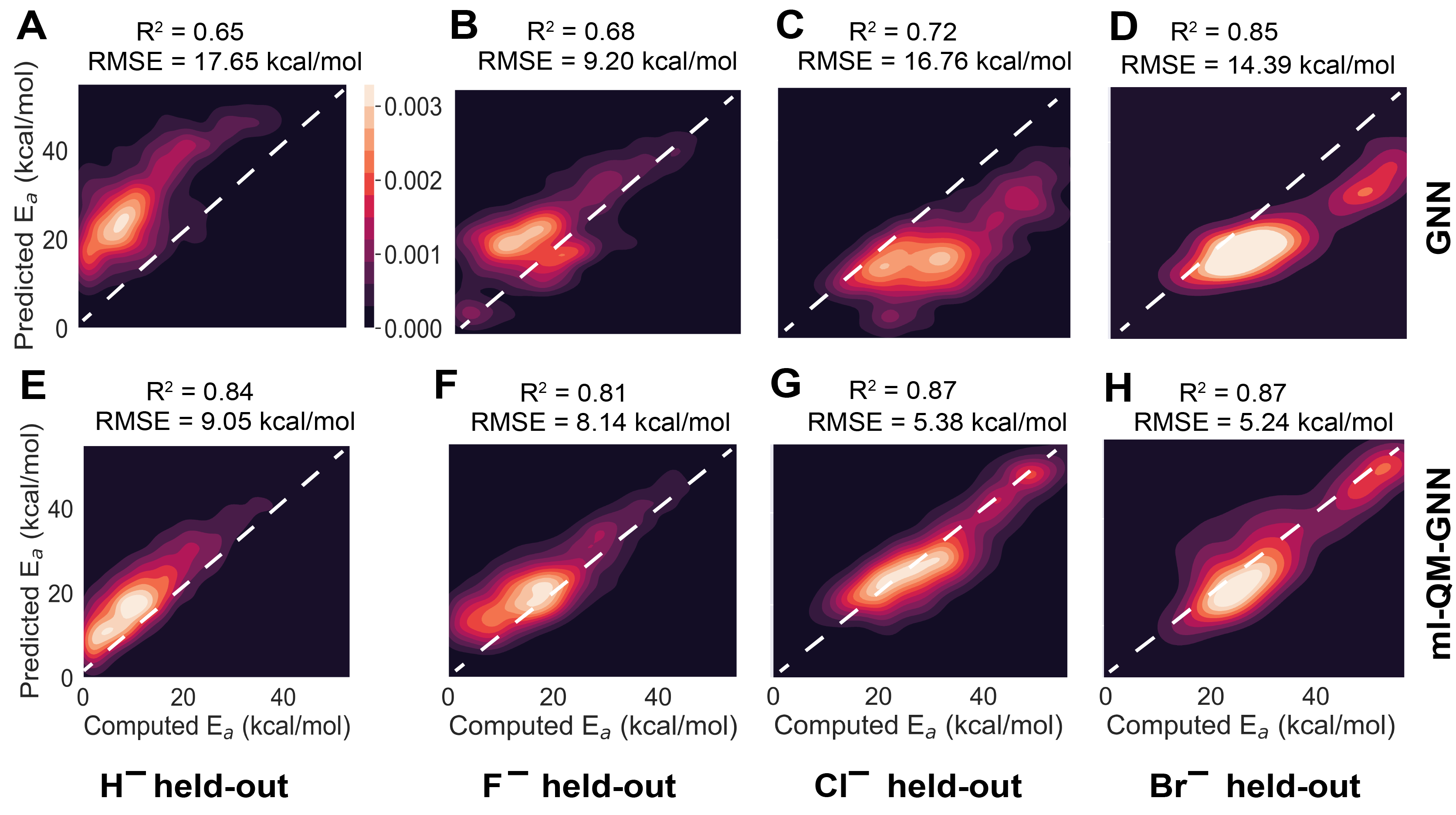}
\caption{Correlation plots for the aggregated predictions made by the regular GNN model across iterations, with held-out nucleophiles (a) H$^-$, (b) F$^-$, (c) Cl$^-$ and (d) Br$^-$. Correlation plots for the ml-QM-GNN model with held-out nucleophiles (e) H$^-$, (f) F$^-$, (g) Cl$^-$ and (h) Br$^-$. The mean RMSE is shown at the top of each individual panel (cf. Section \ref{sec:stdev} for the obtained standard deviations).}
\label{fig:selective_sampling}
\end{figure}

To assess model generalizability, training/validation and test sets were selectively sampled so that for one of the four nucleophiles in the data set, all its reactions would consistently be part of the test set, and consequently this nucleophile is not "seen" by the model during training. The random train and validation set sampling (in a 3:1 ratio) was iterated 5 times and the resulting predictions were aggregated. In Figure \ref{fig:selective_sampling}, correlation plots for each of the different "held-out" nucleophiles are presented. 

The regular GNN models barely manage to reproduce the qualitative trend in the computed activation energies; they significantly underestimate the quantitative values for the barriers when either the hydride or fluoride nucleophiles are held-out, whereas the barriers are vastly underestimated when chloride and bromide are held-out during training (mean RMSEs range from 9 to 18 kcal/mol). KRR in combination with one-hot encoding \cite{heinen2020quantum} performs even worse than the regular GNN on this task; upon selectively sampling, RMSEs between 9 and 20 kcal/mol are obtained for this model architecture (cf. Table \ref{tbl:one-hot-encoding} in the Supporting Information). The ml-QM-GNN on the other hand obtains decent correlations for each of the models trained with hold-out nucleophiles (Figure \ref{fig:selective_sampling}e-h), and the quantitative agreement between model predictions and true values is much better: for hydride and fluoride, the mean RMSEs amount to 8-9 kcal/mol, whereas for chloride and bromide, the mean RMSEs amount to a reasonable 5-6 kcal/mol. 

These results constitute an unequivocal demonstration that designing the GNN model to construct a QM-based representation before the final reactivity prediction not only improves the model accuracy in this data-limited setting, but also improves the model's ability to generalize to unseen nucleophiles.

\subsection*{Interpretability -- E2/S$_\text{N}$2 data set}

Finally, we aimed to gain some insights into how the model reaches its decisions/reactivity predictions. It is not always apparent what exactly deep learning models are learning and how they generalize to new, previously unseen data points, which makes their performance less predictable in prospective settings. We performed a set of ablation experiments, where we controlled the number and type of atom-centered QM descriptors that are used to supplement the structural representation. Specifically, we masked either the nucleophilic and electrophilic Fukui indices or the Hirshfeld partial charges and NMR shielding constants.  

The decision to consider the effect of these pairs of descriptors simultaneously is inspired by the core principle emerging from physical organic chemistry/cDFT that the interactions between reacting species can generally be subdivided in two main types: "hard-hard" or electrostatic interactions, and "soft-soft" or (frontier) orbital interactions. \cite{klopman1968chemical,pearson1963hard} Fukui functions, respectively defined as the (atom-condensed) distribution of an added and removed electron to the system, probe the latter, whereas atomic charges and NMR shielding constants, which reflect the (de)shielding of the nuclei caused by intramolecular electron donation/withdrawal, probe the former. \cite{geerlings2003conceptual, stuyver2020unifying, anderson2007conceptual}

Table \ref{tbl:sn2_e2_preference} contains the accuracies obtained after three 5-fold CVs for the baseline GNN model and the models with one or both sets of estimated QM descriptors. For the baseline model, an average RMSE of 10.1 kcal/mol was obtained; for the full QM-augmented model (``ml-QM-GNN (full)''), the average RMSE was reduced to 3.9 kcal/mol. Remarkably, the ablated model that bases its reactivity predictions solely on atomic charges and NMR shielding constants as QM descriptors (``ml-QM-GNN (charge + NMR)'') recovers the exact same accuracy as the full QM-augmented model (RMSE = 3.9 kcal/mol); the ablated model that only makes use of the Fukui functions (``ml-QM-GNN (Fukui)'') on the other hand achieves the same accuracy as the baseline model (RMSE = 10.0 kcal/mol). These findings suggest that electrostatic/hard-hard interactions are the main drivers of the variations observed in the data set and that Fukui functions, i.e., frontier orbital interactions, are not particularly relevant in this regard.

Further evidence that Fukui functions do not play a significant role in the decision-making process of our QM-augmented network is obtained when the fully trained QM-augmented GNN model is applied to all data points after averaging Fukui function values across all atoms. In this setup, the model predictions are barely affected ($R^2 = 0.96$ between the original predictions and the predictions with Fukui information averaged out). Averaging out the charges and NMR shielding constants on the other hand scrambles the predictions made by the model entirely (Figure \ref{fig:scrambled}).

The same trends emerge for the classification models constructed for the second curated data set, also extracted from the E2/S$_\text{N}2$ reaction data (Table \ref{tbl:sn2_e2_preference}). Again, an improved accuracy is observed when the model bases its predictions on the full set of predicted QM descriptors (77\% to 89\% accuracy). Removing the Fukui function information from the QM-augmented model does not affect the accuracy in a meaningful way, whereas removing the charge and NMR shielding constant data again causes the accuracy to drop to the baseline level.

\begin{table}
  \caption{Model performance as a function of QM descriptor set inclusion. Mean RMSEs and standard deviations (kcal/mol) obtained for the E2/S$_\text{N}2$ barrier height prediction from three random 5-fold cross-validations for the different (ablated) GNN models tested, as well as the corresponding classification accuracies (and their standard deviations) for the prediction of E2 versus S$_\text{N}2$ preference. Standard deviations were again determined from the three replicates.}
  \label{tbl:sn2_e2_preference}
  \begin{tabular}{lcc}
    \hline
    model &  RMSE (kcal/mol) & accuracy  \\
    \hline
    baseline GNN & $10.13 \pm 0.14$ & ($77.0 \pm 0.8$) \% \\
    ml-QM-GNN (full) & $3.92 \pm 0.01$ & ($89.0 \pm 0.6$) \% \\
    ml-QM-GNN (Fukui) & $10.02 \pm 0.02$ & ($77.1 \pm 0.1$) \% \\
    ml-QM-GNN (charge + NMR) & $3.93 \pm 0.02$ & ($88.9 \pm 0.6$) \% \\
    \hline
  \end{tabular}
\end{table}

\begin{figure}[h]
\centering
\includegraphics[scale=0.6]{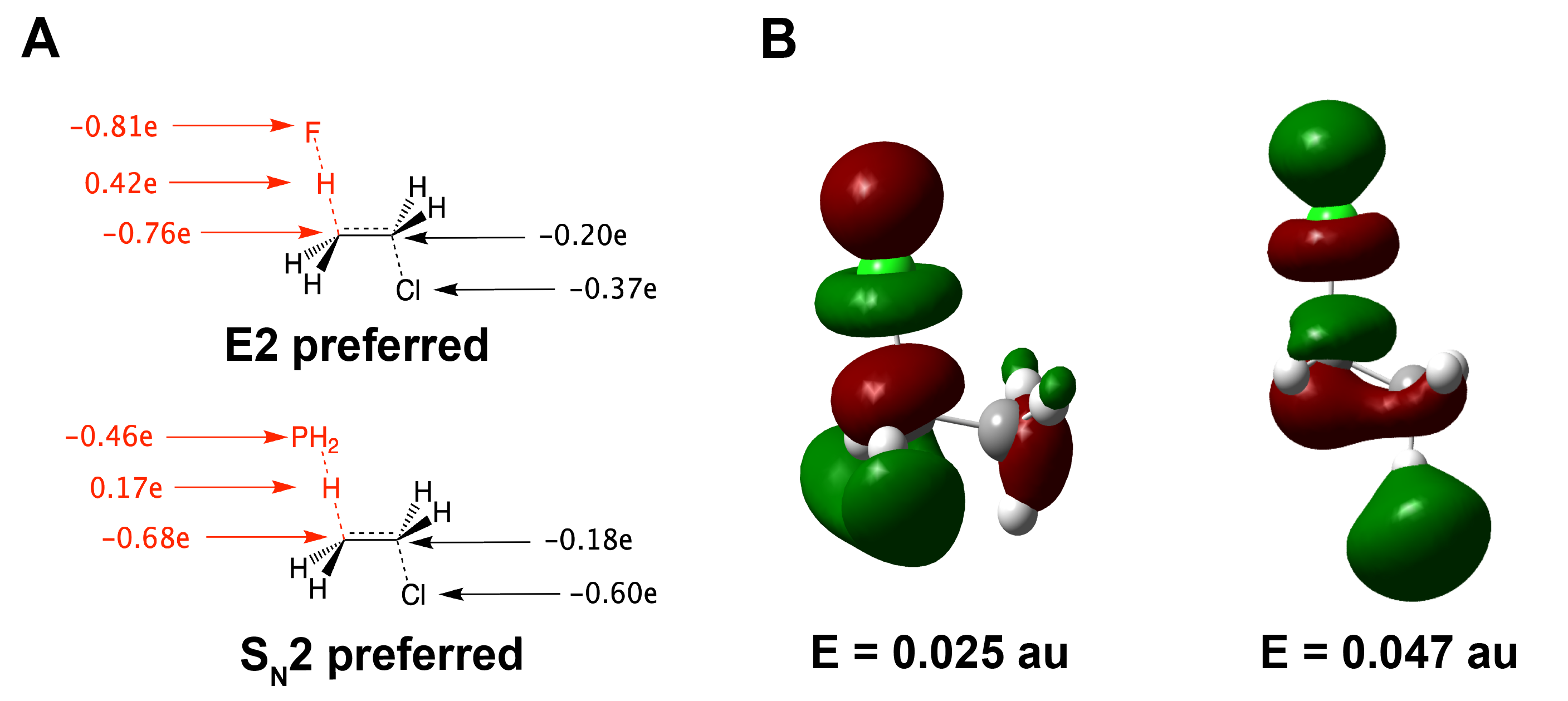}
\caption{(a) Partial (NPA) charge distribution in the E2-TS geometry for F$^-$ + H$_3$CCH$_2$Cl (top) and H$_2$P$^-$ + H$_3$CCH$_2$Cl (bottom); two model systems frequently used as a starting point for qualitative analyses of E2/S$_\text{N}$2 competition. The more pronounced (-)--(+)--(-) charge array in the case of the F$^-$ nucleophile causes sufficient electrostatic stabilization for the E2-TS to drop below the S$_\text{N}$2-TS in energy; in the case of the  H$_2$P$^-$ nucleophile, the the reduced electrostatic stabilization causes the S$_\text{N}$2-TS to remain lower in energy. (b) The lowest unoccupied molecular orbital (LUMO; E = 0.025 au; left) and LUMO+1 (E = 0.047 au; right) for H$_3$CCH$_2$Cl. Calculations were performed at M06/def2-TZVP level-of-theory (cf. Section \ref{sec:dft_methodology}). \cite{stuyver2021resolving}}
\label{conceptual_interpretation}
\end{figure}

Our findings about the relative importance of electrostatic/"hard-hard" vs. (frontier) orbital/"soft-soft" interactions, emerging in both regression and classification tasks, are perfectly in line with a recent qualitative VB/cDFT analysis undertaken by one of the authors of the present work. \cite{stuyver2021resolving} In this analysis, it was demonstrated that the modulation of the E2/S$_\text{N}$2 competition is primarily driven by the electrostatic interactions present in the E2-TS: the formation of a strongly stabilizing array of point-charges, i.e., (-)--(+)--(-), in this geometry tends to push its energy below that of the S$_\text{N}$2-TS; in the case that the point-charges in this array -- and thus the Coulombic interaction -- are weaker, the S$_\text{N}$2-pathway dominates (cf. Figure \ref{conceptual_interpretation}a).

The observation that Fukui function values are not helpful to a GNN aiming to learn this mechanistic competition could also have been readily anticipated from the analysis presented in the same qualitative study: the ethylhalide substrates on which the nucleophiles attack generally carry two close-lying, unoccupied frontier orbitals which are delocalized over both the $\alpha$-carbon and the hydrogen on the $\beta$-carbon (Figure~\ref{conceptual_interpretation}b) \cite{stuyver2021resolving};  providing information about only the lowest-lying of these orbitals through the (electrophilic) Fukui function is not informative at all and essentially constitutes noise. 
The discussion above demonstrates that combining a qualitative analysis rooted in conceptual reactivity frameworks with our QM-augmented machine learning approach leads to a productive synergy for the considered data set of competing E2/S$_\text{N}$2 reactions: on the one hand, the qualitative insights provide context to interpret and understand the decision/prediction-making process of the network. At the same time, the results emerging from our ablation study can also be considered as an indirect, data-driven confirmation of the qualitative reactivity analysis.

\subsection*{Interpreting the predicted regioselectivity of aromatic substitution reactions}

To further explore the interpretability of ml-QM-GNN models, we revisited the classification data set containing regiochemical data for electrophilic substitution reactions, originally compiled by Guan et al. to demonstrate the potential of this model architecture in data-limited settings. \cite{guan2021regio} As mentioned in the Computational Details section,  we select only those data points which are publicly available in the USPTO database, resulting in a curated data set of 3242 reactions. In Table \ref{tbl:aromatic_preference}, the accuracies achieved by the baseline GNN, the ml-QM-GNN, as well as the two ablated models on the curated data set (training set limited to 200 data points) are presented.
\begin{table}
  \caption{Average accuracies (and their standard deviation across replicates) obtained from three random 5-fold CVs for the different (ablated) GNN models applied to the curated aromatic substitution data set (training set limited to 200 data points).}
  \label{tbl:aromatic_preference}
  \begin{tabular}{lc}
    \hline
    model & accuracy   \\
    \hline
    baseline GNN & $71.4 \pm 1.3$ \% \\
    ml-QM-GNN (full) & $86.5 \pm 0.7$ \% \\
    ml-QM-GNN (Fukui) & $84.4 \pm 1.1$ \% \\
    ml-QM-GNN (charge + NMR) & $82.2 \pm 1.3$ \% \\
    \hline
  \end{tabular}
\end{table}
Once more, we observe that the full QM-augmented model significantly outperforms the baseline GNN. More interestingly, it can be observed now that \emph{each} set of descriptors leads to a significant improvement in accuracy over the baseline but that \emph{both} are needed to maximize performance.  

Even though the differences are rather small, the results presented in Table \ref{tbl:aromatic_preference} suggest that for this class of reactions, the Fukui functions are slightly more informative than charges and NMR shielding constants, which is in contrast to what was observed for the E2/S$_\text{N}$2 data set in the previous section. 
These findings can be straightforwardly reconciled with the qualitative physical organic reactivity models that have been constructed and popularized throughout the years: aromatic compounds are strongly delocalized, and hence, orbital interactions/changes in delocalization stabilization as probed through the Fukui functions are generally considered to be the main driving force shaping their reactivity. \cite{fukui1954molecular, langenaeker1991quantum,stuyver2020local} At the same time, it has been underscored in recent years that electrostatics can not be neglected either---in particular under conditions favoring kinetic control---since atomic charges tend to become more pronounced as the compounds involved in the reaction approach, \cite{shaik1999valence} enabling a significant Coulombic stabilization/destabilization of the wave function in the transition state region. \cite{stuyver2020unifying, anderson2007conceptual}

\begin{figure}[h]
\centering
\includegraphics[width=3.4in]{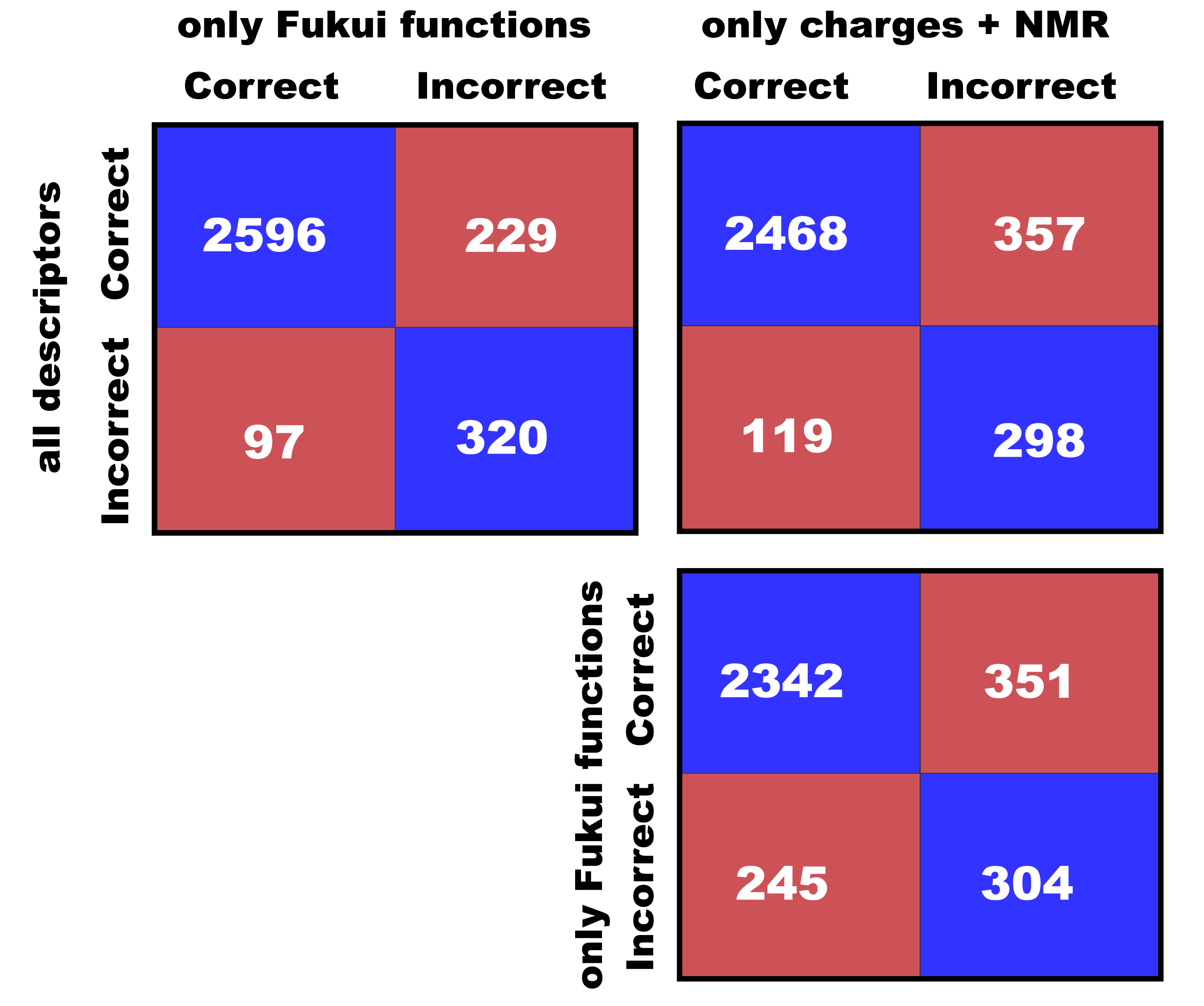}
\caption{Confusion matrices comparing the predictions made by the different (ablated) ml-QM-GNN models for all test sets sampled during the first 5-fold CV combined. The labels in the margins of the individual matrices indicate which descriptors are considered by the respective model, i.e., either only the soft-soft or hard-hard descriptors or both descriptor-types combined.}
\label{fig:confusion_matrices}
\end{figure}

To obtain a better understanding of how the (ablated) QM-augmented models reach their decisions/predictions, we constructed a set of confusion matrices comparing the respective prediction accuracies for all test sets considered during the first cross-validation (Figure \ref{fig:confusion_matrices}). 
These matrices show that there are relatively few data points (97 and 119) where an ablated models makes the correct prediction while the model with access to all descriptors reaches the wrong conclusion; the full QM-augmented model appears to mainly rectify incorrect predictions by the ablated models (cf. the upper two square matrices in Figure \ref{fig:confusion_matrices}). Additionally,  the lower-right confusion matrix in Figure \ref{fig:confusion_matrices} demonstrates that many incorrect predictions by the two ablated models are distinct: the number of points for which there is disagreement between the models (245 + 351), is greater than the off-diagonal elements in the upper matrices.

We then considered whether incorrect prediction of regioselectivity preference is connected to failures of either the "hard-hard"/electrostatic or "soft-soft"/(frontier) orbital criterion. The "hard-hard"/electrostatic criterion is considered to be fulfilled  when the reacting site on the aromatic substrate corresponds to the site carrying the highest partial negative charge (in the case of electrophilic attack) or partial positive charge (in the case of nucleophilic attack), since this would maximize the Coulombic stabilization upon approach between the reacting species. The "soft-soft"/(frontier) orbital criterion is considered to be fulfilled when the reacting site on the substrate corresponds to the site on which the nucleophilic Fukui function is most concentrated (in the case of electrophilic attack) or the site on which the electrophilic Fukui function is most concentrated (in the case of nucleophilic attack) -- which would maximize the orbital interaction upon approach between the reacting species. 

\begin{figure}[h]
\centering
\includegraphics[scale=0.6]{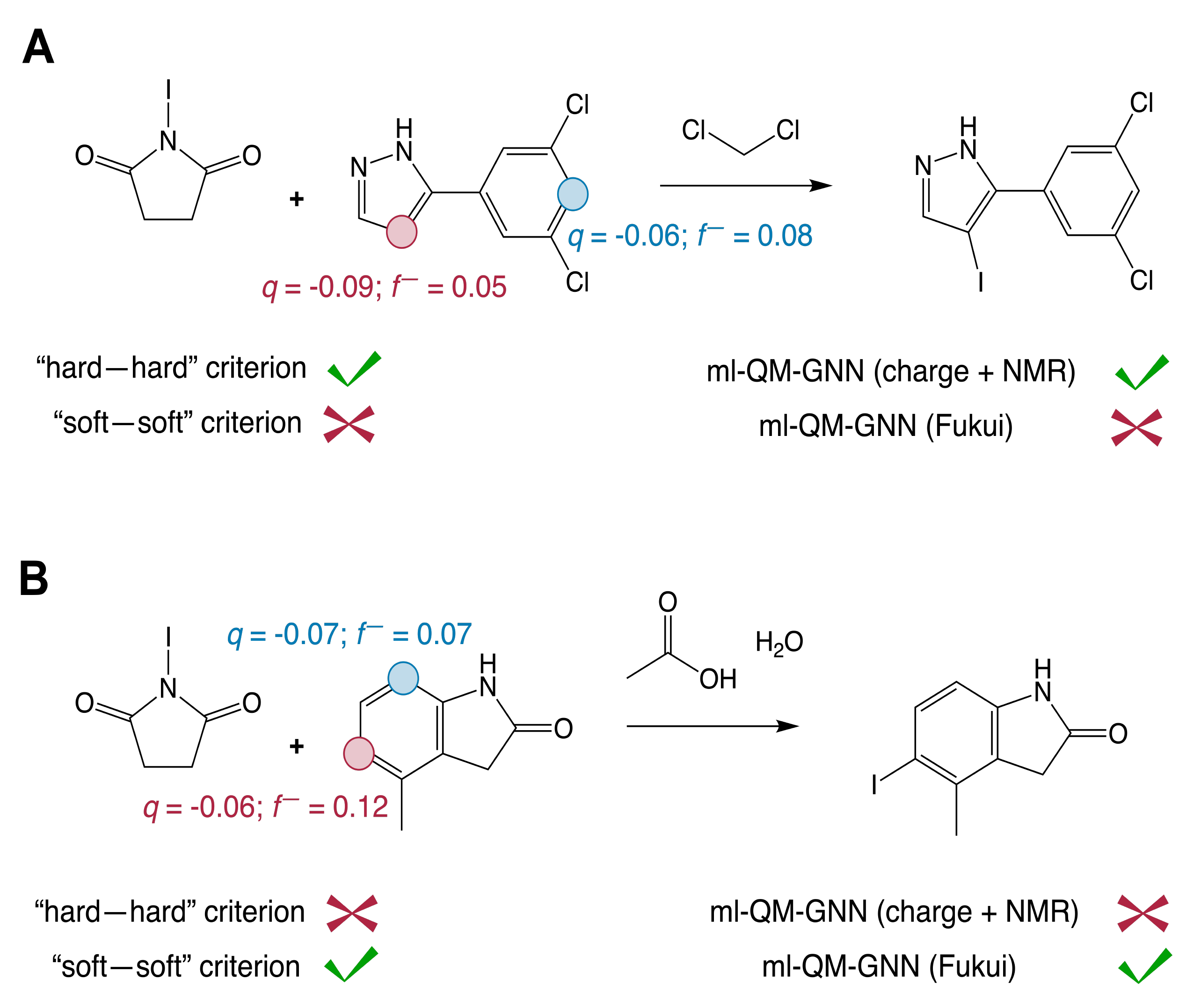}
\caption{The predictions made by the ablated models generally adhere to their corresponding physical organic criterion ($q$ denotes the predicted partial charge; $f^-$ denotes the predicted  atom-condensed nucleophilic Fukui function value; the red dots correspond to the reported/"true" reactive sites, the blue dots are competing sites). (a) For this example reaction, the hard-hard criterion is fulfilled, but the soft-soft one is not; correspondingly, the ablated model which considers exclusively the partial charges an NMR shielding constants makes the correct prediction and the ablated model which considers exclusively Fukui function values makes the incorrect one. (b) An example reaction for which the situation is reversed.}
\label{fig:ablated_confusion2}
\end{figure}

Since nitration reactions and halogenation reactions involving N-substituted succinimides are electrophilic without exception, are straightforward to recognize based on a SMILES representation, and collectively cover approximately half of the data set (1534 out of 3242 data points), we decided to focus exclusively on these reaction types for this part of our analysis. A confusion matrix comparing the predictions obtained through naive evaluation of the individual physical organic criteria (\emph{vide supra}) with the predictions made by the (ablated) models during 5-fold CV is presented in Table \ref{tbl:testset_concatenated}. The predictions made by the models tend to adhere to the respective criteria: the vast majority of data points for which both criteria point to the correct reactive site are classified correctly by the ablated and the full QM-augmented GNNs, whereas reactions incorrectly classified only by the ablated model that exclusively considers Fukui function values disproportionately violate the "soft-soft" criterion (and vice versa: the reactions that are only incorrectly classified by the ml-QM-GNN (charge + NMR) disproportionately violate the "hard-hard" criterion; Figure~\ref{fig:ablated_confusion2}). Only a handful of reactions for which neither criterion is fulfilled are classified correctly.

Putting all this together, one can conclude that the trained QM-augmented GNN model with all descriptors included appears to be able to balance the relative importance of the electrostatic and (frontier) orbital criteria, whereas the ablated models are generally biased toward (a) assigning too much importance to the interaction type which they have access to, and (b) adhering to the rudimentary assignment based on the "soft-soft"/"hard-hard" criterion, even when this assignment is incorrect. 

As such, our ml-QM-GNN model essentially refines the balancing act that is implicitly part of any cDFT-based qualitative reactivity analysis. As indicated above, most human theoreticians tend to assume that the regioselectivity of aromatic reactions is primarily determined by the Fukui functions, i.e., the soft-soft interactions dominate, and they will often ignore the electrostatics altogether (or assume that these are in sync with the Fukui function values). \cite{langenaeker1991quantum} Our analysis (Section \ref{sec:confusion}) shows that this is not an unreasonable approximation to make for this dataset;  selecting the site to which the soft-soft criterion points---ignoring the hard-hard criterion completely---results in an accuracy of $\approx$80\%. By relaxing this rigid guiding rule and replacing it by a more complex classification function centered around the same QM descriptors, our models reach accuracies exceeding 86\% when a mere 200 training points are used (Table \ref{tbl:aromatic_preference}), and reaching up to 93-94\% when the training set size is expanded to 1945 training points.

\subsection*{Failures in predicting regioselectivity of aromatic substitution reactions}

Finally, we also took a closer look at some of the reacting systems for which neither criterion is fulfilled and for which all our GNN models fail to make the correct prediction. While it appears impossible to assign a specific reason for each failure, there are some recurring patterns.

\begin{figure}[h]
\centering
\includegraphics[scale=0.57]{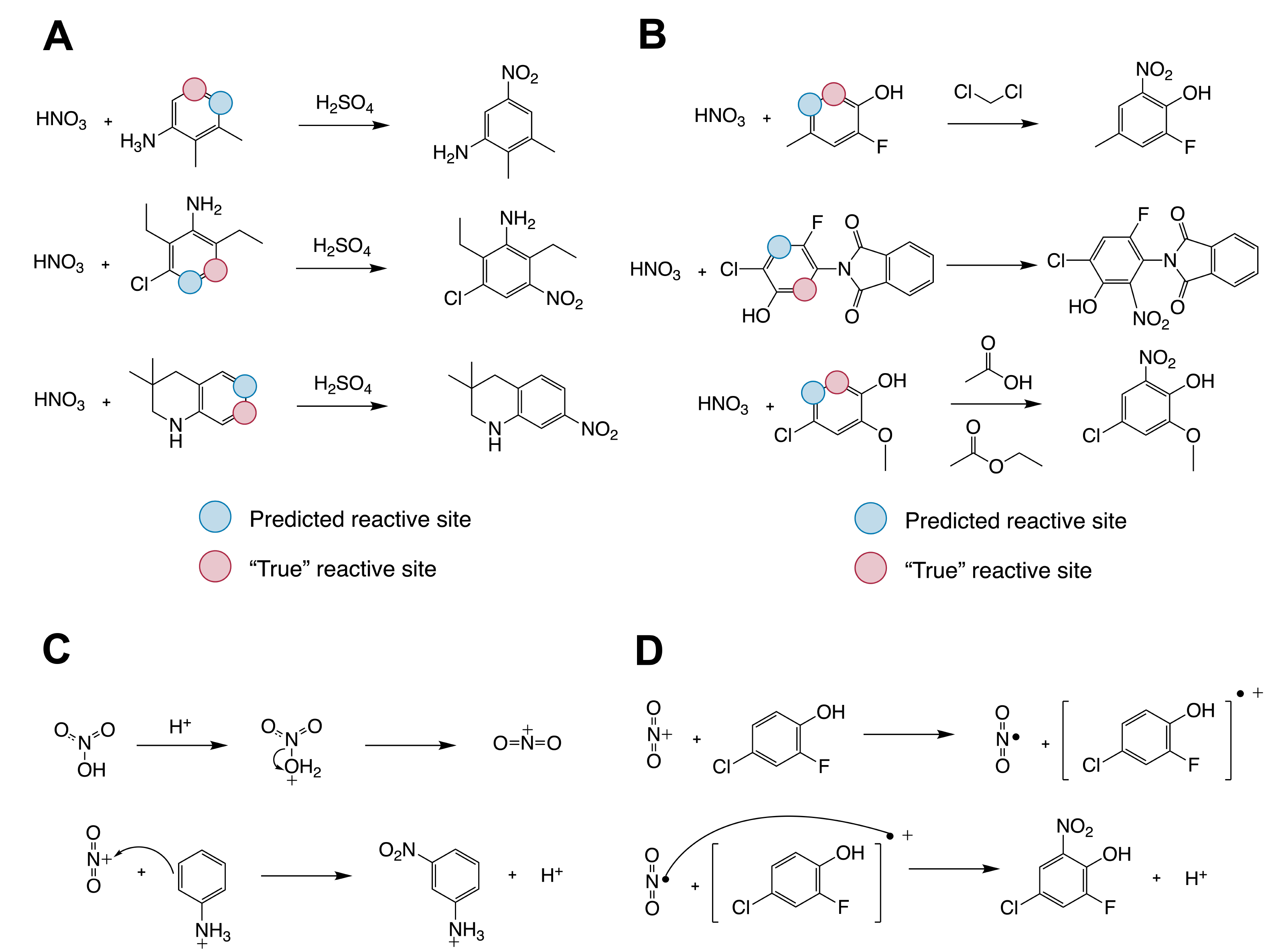}
\caption{(a) Some examples of nitration reactions with an NH$_2$-substituent on the aromatic substrate for which both physical organic criteria and all QM-augmented models fail (the blue dots indicate the -- incorrectly -- predicted sites; the red dots indicate the true reactive sites). (b) Some examples of nitration reactions with an OH-substituent on the aromatic substrate for which both physical organic criteria and all QM-augmented models fail. (c) The mechanism of a regular (electrophilic) nitration reaction involving a (protonated) aniline substrate. (d) The proposed radical mechanism for nitration of halogenated phenol analogues. \cite{ducry2005controlled}}
\label{fig:failures}
\end{figure}

As an example, all criteria and models consistently appear to fail for most nitration reactions involving aniline analogues, i.e., aromatic substrates with an NH$_2$-substituent (Figure \ref{fig:failures}a). For this type of reaction, nitration in ortho- or para-position is expected, but the "true" reactive site corresponds to the meta-position. This failure is a reflection of the fact that nitration reactions are usually performed in strongly acidic reaction media to promote the protonation of nitric acid into the active nitronium ion species, which inherently results in simultaneous protonation of the NH$_2$-substituent, rendering this substituent meta-directing instead of para-directing (Figure \ref{fig:failures}c). 

Another recurring failure is observed for nitration reactions involving halogenated phenols. In this type of reaction, both the criteria and models predict the reaction to occur in ortho-/para-position relative to the halogen substituent, whereas the "true"/recorded reactive site is consistently the ortho-position relative to the OH-substituent (Figure \ref{fig:failures}b). This failure can most likely be attributed to a mechanistic cross-over: instead of a conventional electrophilic aromatic substitution mechanism, reactions between phenolic compounds and nitric acid/nitronium ion have been demonstrated to involve a (pure) single electron transfer step, followed by radical recombination (Figure \ref{fig:failures}d). \cite{ducry2005controlled, perrin1977necessity} This change in mechanism can be expected to impact the balance in directing strengths of the halogen- and hydroxy-substituents with respect to regular electrophilic aromatic substitution reactions, resulting in a modification of the regiochemistry.

Since both the physical organic criteria and our GNN models are agnostic to reaction conditions and "concealed" reaction steps, they are unable to capture these modifications to the directing character of the NH$_2$ and OH-substituents. One can anticipate that by adding more and more training data, the model may learn to overrule the "regular" QM descriptor assignment when NH$_2$/OH-substituents and nitric acid are simultaneously present in the reacting system, but since only 200 data points have been used for training here, there are simply not enough examples to capture these particular patterns.

\section*{Conclusions}

In this work, the performance, interpretability and generalizability of the QM-augmented GNN (ml-QM-GNN) model architecture has been assessed for a few distinct predictive chemistry tasks. Our models achieve a significantly improved accuracy over a conventional GNN baseline and generalize markedly better to unseen compounds, particularly in a data-limited regime. Even when only a couple hundred labeled data points are available, our ml-QM-GNN models are competitive with traditional ``low-data'' KRR models. 

Importantly, since the predictions made by our models are rooted in (predicted) QM descriptors, it becomes possible to build bridges between their predictions and the existing physical organic frameworks, developed to qualitatively analyze chemical reactivity. 
Through a series of selective descriptor ablation experiments, we have demonstrated that for competing E2/S$_\text{N}$2 reactions, Fukui function values, i.e., information about soft-soft interactions, are of limited value; charges and NMR shielding constants are the main drivers of the improvement in model accuracy with respect to the baseline GNN model. These findings can be rationalized through consideration of a recent qualitative valence bond/cDFT analysis of this mechanistic competition. \cite{stuyver2021resolving} 

For aromatic substitution reactions, we observed that our ml-QM-GNN model appears to make its decision/predictions in a similar manner as human theoreticians  trained in physical organic chemistry/cDFT, i.e., by considering soft-soft and hard-hard interactions separately through their corresponding local descriptors. What makes our models excel with respect to a naive conceptual treatment is their ability to fine-tune the relative importance of the individual physical organic criteria based on subtle patterns present in the data, resulting a more complex classification function and a significantly higher accuracy.

Overall, our analysis underscores that a productive interplay between machine learning models and qualitative reactivity analysis is possible: on the one hand, qualitative insights into the considered reactivity problem provides context to interpret and understand the decision-making process of the network. Additionally, they can provide clues about the suitability to include specific QM descriptors in the neural network.  At the same time, the results emerging from machine learning models augmented with QM descriptor information can provide an indirect, data-driven confirmation of a qualitative reactivity analysis.

\begin{acknowledgement}

The authors thank the Machine Learning for Pharmaceutical Discovery and Synthesis (MLPDS) consortium for funding and Yanfei Guan for helpful discussions.

\end{acknowledgement}

\begin{suppinfo}

In-depth technical description of the GNN model architectures, technical details related to the overall training process and cross-validation, the effect of removing "duplicate" data points from the E2/S$_\text{N}$2 data set, note regarding the data curation procedure for the E2/S$_\text{N}$2 classification data set, selective sampling results for one-hot encoding in combination with KRR, RMSEs and standard deviations obtained for the GNN and ml-QM-GNN during selective sampling, the effect of averaging out the respective descriptors for the full QM-augmented model trained on the E2/S$_\text{N}$2 data set, Methodology for the DFT calculations in Figure \ref{conceptual_interpretation}, confusion matrix comparing the classifications by the QM-augmented models to the physical organic criteria for the aromatic substitution data set. The main code and data sets used in this work can be found at \url{https://github.com/coleygroup/QM-augmented_GNN}.

\end{suppinfo}

\putbib[biblio_ms]
\end{bibunit}

\clearpage
\newpage

\begin{bibunit}[unsrt]

\setcounter{section}{0}
\setcounter{equation}{0}
\setcounter{figure}{0}
\setcounter{table}{0}
\section*{Supplementary Information}
\normalsize
\setcounter{section}{0}
\renewcommand\thesection{S\arabic{section}}
\def\thefigure{S\arabic{figure}}
\def\thetable{S\arabic{table}}
\def\theequation{S\arabic{equation}}

\section{In-depth technical description of the GNN model architectures}\label{sec:architecture}

\subsection{The baseline GNN model}

When a new data point is fed to the baseline GNN model, the first step that is taken is a conversion of the SMILES representation of the reactants to a graph-based/structural one. More specifically, each atom $v$ is initialized with a feature vector $f_v$ indicating its atomic number, degree of connectivity, explicit and implicit valence, and aromaticity. Each bond (u, v) is associated with a feature vector $f_{uv}$ indicating its bond type and ring status. Subsequently, the structural representation is fed to a Weisfeiler-Lehman network (WLN). \cite{NIPS2017_ced556cd} This network constructs a convolutional embedding of the atoms by iteratively updating their respective structural representation with information from adjacent atoms. In the $t^{\text{th}}$ iteration, the atom representation is updated from $f_v^{t-1}$ to $f_v^t$ as follows,
\begin{equation}
f_v^t = \text{ReLU}(U_1(V_1 f_v^{t-1} \odot \sum_{u \epsilon N(v)} \text{ReLU}(U_2(W_1 f_v^{t-1} \odot W_2 f_uv))))
\end{equation}
where ReLU is the rectified linear unit, $\odot$ corresponds to the Hadamard product and U$_i$, V$_i$ and W$_i$ are learned matrices. After L iterations (in our architecture, WL networks of 4 consecutive layers were used consistently), the final local atomic representations are computed as follows,
\begin{equation}
c_v = V_1 f_v^{L} \otimes \sum_{u \epsilon N(v)} W_1 f_u^{L}
\end{equation}
where $\otimes$ denotes the tensor product.

The atomic embeddings $c_v$ only encode local structural patterns, namely atoms and bonds accessible within $L$ steps from atom $v$. To capture distant information (e.g., information between disconnected atoms), the resulting embeddings $c_v$ are then passed through an attention layer, which calculates the so-called "attention score" of atom $v$ on atom $z$. Attention scores for atom pair ($v$,$z$), $\alpha_{vz}$, are calculated by this layer as follows,
\begin{equation}
\alpha_{vz} = \sigma(Q_1 \text{ReLU}(P_1(c_v + c_z) + P_2 b_{vz})
\end{equation}
where $\sigma$ indicates a sigmoid activation function and $P_1$, $P_2$ and $Q_1$ are learned matrices. The "global" atom representation $\tilde{c}_v$ is defined as the weighted sum of all reactant atom feature vectors,
\begin{equation}
\tilde{c}_v = \sum_z \alpha_{vz} c_z
\end{equation}
The final atomic representation for atom $v$, $\hat{c}_v$, is then constructed as follows,
\begin{equation}
\hat{c}_v = \text{ReLU}(M(\tilde{c}_v + c_v))
\end{equation}
where M is the learned matrix. This final atomic representation is then passed through a sum-pooling layer, which sums over the atomic representations of the atoms which undergo a change in bonding situation throughout the reaction, i.e., the atoms part of the reacting core ($RC$), resulting in the reaction representation. From this reaction representation, a final reaction score is determined by passing the reaction representation through a final feed forward layer. In the case of a regression task, the reaction score can be expressed as follows,
\begin{equation}
s = O(\sum_{u \epsilon RC} \hat{c}_v)
\end{equation}
with $O$ a learned matrix and $s$ is the reaction score. In the case of classification, a \emph{softmax} activation function, $\tau$, is included in this final layer as well, i.e.,
\begin{equation}
s = \tau(O(\sum_{u \epsilon RC} \hat{c}_v))
\end{equation}

\subsection{The ml-QM-GNN model}

As mentioned in the main text, the QM-augmented GNN model follows a similar architecture as the baseline GNN model, with the main difference being the presence of a QM descriptor prediction module (the D-MPNN branch, \emph{vide infra}), which branches off immediately following the construction of the graph-based input representation. 

As indicated in the main text, the predicted descriptors can be subdivided in atom-centered and bond descriptors. The atom-centered descriptors consist of partial charges, and electrophilic/nucleophilic Fukui functions (the Hirshfeld population analysis scheme \cite{hirshfeld1977bonded} was consistently applied for each of these), as well as NMR shielding constants (as computed with the help of the Gauge-Independent Atomic Orbital (GIAO) method). \cite{wolinski1990efficient} The bond descriptors are the bond lengths, determined from the GFN2-xtb-optimized geometries, \cite{bannwarth2019gfn2} and the NPA bond orders, determined with NBO 6.0. \cite{glendening2013nbo}

The atomic descriptors are normalized with the help of min-max normalization. For the NMR shielding constants, the scaling was performed on an element-by-element basis due to the dramatic differences in shielding constant magnitudes depending on the atom-type, which completely dwarf the typical intra-element variation. For similar reasons, the anionic nucleophiles in the E2/S$_\text{N}$2 data set \cite{von2020thousands} (H$^-$, F$^-$, Cl$^-$ and Br$^-$) were not taken into account during the parametrization of partial charge and NMR shielding constant min-max scaler, since the corresponding values carried by those anions are so extreme that they would "compress" the regular variation observed for these atomic descriptors if they were included (\emph{vide infra}).  

All continuous descriptors outputted by the D-MPNN branch are turned into vector representations with the help of a radial basis function (RBF) expansion. In the case of continuous bond descriptors, $b_{uv}$, the vector representation, $e_{uv}$ can be expressed as,
\begin{equation}
e_{uv} = [\exp{(-\frac{(b_{uv} - (\mu + \delta k))^2}{\delta})}]_{k\epsilon[0,1,2,...,n]}
\end{equation}
where $n$ is the number of basis function -- and thus also the length of the vector representation -- and $\mu$ and $\delta$ are the parameters which determine the starting point and resolution of the expansion respectively. For example, for bond lengths, which typically range from 0.5 to 2.5 \AA, $\mu$, $\delta$ and $n$ are chosen to be 0.5, 0.05 and 40 respectively, which expands the bond length into a vector of size 40 with an interval of 0.05 \AA, starting from 0.5 \AA. 

The initial bond vector $f_{uv}$, which is immediately fed to the WL branch of the network, is computed as,
\begin{equation}
f_{uv} = e_{uv}^{bl} \oplus e_{uv}^{bo}
\end{equation}
where $\oplus$ stands for vector concatenation and $e_{uv}^{bl}$ and $e_{uv}^{bo}$ are the expanded vectors for bond length and bond order respectively.

In the case of the continuous atom-centered descriptors, $a_u$, the vector representation $e_u$ is expressed as follows,
\begin{equation}
e_{u} = [\exp{(-\frac{(a_{u} - (\mu + \delta k))^2}{\delta})}]_{k\epsilon[0,1,2,...,n]}
\end{equation}
Since in the E2/S$_\text{N}$2 data set, the scaled descriptor values for the partial charge and NMR shielding constant on the nucleophiles inherently are not confined to [0,1] (\emph{vide supra}), the respective RBFs were extended beyond this range so that these outlier values could be covered in the vector representation without compressing the variation among the non-outlier values on a small number of vector elements. This is done to maintain a comparable resolution of the descriptor variation for all elements on the regular (non-anionic) compounds across the different considered data sets, since this is vital for the model to be able to exploit the chemically relevant patterns and trends encoded in these variations. In Table \ref{tbl:RBF parameters}, an overview of the respective $\mu$, $\delta$ and $n$ parameters selected for each descriptor is presented.  

\begin{table}
  \caption{Parameters for the RBF expansion for each descriptor.}
  \label{tbl:RBF parameters}
  \begin{tabular}{cccc}
    \hline
    descriptor & $\mu$ & $\delta$ & $n$ \\
    \hline
    partial charge ($q$) & -2.0 & 0.06 & 50 \\
    electrophilic Fukui function ($f^+$) & 0.0 & 0.02 & 50 \\
    nucleophilic Fukui function ($f^-$) & 0.0 & 0.02 & 50 \\
    NMR shielding constant ($sc$) & 0.0 & 0.06 & 50 \\
    bond order ($bo$) & 0.5 & 0.1 & 25 \\
    bond length ($bl$) & 0.5 & 0.05 & 40 \\
    \hline
  \end{tabular}
\end{table}

As indicated in the main-text, the vector representations of the (scaled) atom-centered descriptors are concatenated to the learned atomic-representations emerging from the WL branch, i.e.,
\begin{equation}
\hat{c}_v^{ml-QM-GNN} = \hat{c}_v \oplus a_u^q \oplus a_u^{f+} \oplus a_u^{f-} \oplus a_u^{sc}
\end{equation}
where $a_u^q$ is the partial charge RBF expanded vector, $a_u^{f+}$ and $a_u^{f-}$ are respectively the nucleophilic and electrophilic Fukui function RBF expanded vectors and $a_u^{sc}$ is the shielding constant RBF expanded vector. The resulting concatenated representation is then passed through the attention mechanism and the final dense layer to give the reactivity prediction, in a similar manner as in the baseline GNN (\emph{vide supra}).

\subsection{Architecture of the D-MPNN branch}

The D-MPNN, which encodes the molecular graph into (QM-inspired) atom and bond representations, is an adaptation of ChemProp; \cite{yang2019analyzing} the mathematical background behind the message-passing encoder has been described at length in the original paper in which this model was presented. 

The learned atomic/bond representation emerging from the D-MPNN encoder are converted into the corresponding descriptors through a multi-task readout layer. For unconstrained descriptors, e.g. NMR shielding constants, bond order, and bond length, a simple feed-forward neural network (FFNN) is used to calculate the descriptor from the feature vectors. For constrained descriptors, such as atomic charges and Fukui indices, a regular FFNN first computes the uncorrected descriptors $q_i$ as follows,
\begin{equation}
q_i = \text{FFNN}(\hat{c}_i)
\end{equation}
in which $\hat{c}_i$ denotes the respective atom-centered/bond feature vector. The final corrected descriptor subject to the constraint is then calculated as follows,
\begin{equation}
\hat{a}_i = \text{FFNN}(\hat{c}_i)
\end{equation}
\begin{equation}
w_i = \frac{\exp{u\hat{a}_i}}{\sum_i w_i}
\end{equation}
\begin{equation}
q_i^{final} = q_i + \frac{w_i(Q-\sum_iq_i)}{\sum_i w_i}
\end{equation}
where Q is the constraint applied to the descriptor so that,
\begin{equation}
\sum_iq_i^{final} = Q
\end{equation}

Since the current version of the D-MPNN is only able to treat uncharged compounds, i.e., Q = 0, the descriptor values for the four recurring nucleophiles present in the E2/S$_\text{N}$2 data set, i.e., H$^-$, F$^-$, Cl$^-$ and Br$^-$, were included in the model as fixed parameters.

\subsection{Training of the D-MPNN for reactivity descriptor prediction}

For the training of the reactivity descriptor prediction branch, an aggregate loss function, which aims to minimize the mean squared error (MSE) for each of the considered descriptors simultaneously, was defined (a prefactor has been added to the loss for the NMR shielding constants, to bring the individual terms to approximately the same scale):

\begin{equation}
\text{LOSS} = loss_p + loss_{f^+} + loss_{f^-} + 1e^{-5}*loss_{sc} + loss_{bl} + loss_{bo}
\end{equation}

To generate an exhaustive computational database for training, an automated workflow was originally set up by Guan et al., cf. Ref. \cite{guan2021regio} . This workflow starts by sampling conformers from SMILES strings with the help of the RDKit library, \cite{riniker2015better} and the Merck Molecular Force Field (MMFF94s). \cite{halgren1996merck} The lowest-lying conformer are subsequently optimized at the GFN2-xtb level-of-theory. \cite{bannwarth2019gfn2} A variety of convergence checks are performed to ensure the optimization converged to a correct structure, including checks for imaginary frequencies and ensuring that the molecule does not further converge into other species. The final chemically meaningful descriptors are calculated with the B3LYP functional \cite{becke1993density,lee1988development} in combination with a def2svp basis set. \cite{weigend2005balanced} All DFT computations are performed using Gaussian 16 \cite{Pople2003} (bond orders are calculated with the help of NBO 6.0 \cite{glendening2013nbo}).

In total, 180$k$ organic compounds with a molecular weight below 500 were sampled from the ChEMBL \cite{gaulton2012chembl} and Pistachio \cite{Pistachio} databases and fed to the workflow described above. For the ChEMBL database, 80$k$ molecules containing, C, H, O, N, P, S, F, Cl, Br and I were selected randomly; for the Pistachio database, 100$k$ common reactants, reagents and solvents -- present in more than three reaction records -- were selected. For an overview of the overall performance of the trained D-MPNN, we refer to reference \cite{guan2021regio} .

\section{Technical details related to the overall training process and cross-validation}

The 5-fold cross-validations are performed by first splitting the entire data set into 5 subsets through random sampling. The model is then trained and evaluated 5 times. In each iteration, one subset is picked as the test set, and the validation and training set are randomly sampled from the remaining data. 

In each iteration, the model is trained for maximum 50 epochs. Early stopping based on the loss function on the validating set is employed to prevent over-fitting. The learning rate scheduler is used throughout all models. A reducing learning rate with the decay rate of 0.95 for each epoch has been selected consistently. For the multi-task constrained model, a SINEXP learning rate scheduler as defined in the literature was selected. \cite{wu2019demystifying}

\section{The effect of removing "duplicate" data points from the E2/S$_\text{N}$2 regression data set} \label{sec:duplicate}

During the conversion from reactant complex geometry to SMILES representation, some stereochemical information is lost in the E2/S$_\text{N}2$ data set, resulting in multiple barriers being associated with a single SMILES string, i.e., reacting systems become indistinguishable when prochiral ethylhalide analogues are present in the reacting system. Usually, the activation energies for these "duplicate" data points are very similar, so that one of them can be safely removed in principle. For others however, there is a substantial difference in barrier height, in a couple of extreme cases even exceeding 5 kcal/mol. 

To assess the overall effect of these duplicate data points on the accuracy of the model, we considered what happens when we remove the duplicates, i.e., whenever a SMILES string appeared twice in our curated data set, we removed the highest-lying barrier. As can be noticed from the table below, the RMSEs for the different models considered are largely unaffected by removal of these duplicates, and hence we continued working with all the barriers to facilitate comparison with the KRR models developed by von Lilienfeld and co-workers. \cite{heinen2020quantum}

\begin{table}
  \caption{RMSEs obtained from 5-fold cross-validation for the prediction of the activation energies of the E2 and S$_\text{N}2$ reactions for the different (ablated) GNN models tested, with and without removal of the data points containing identical SMILES representations. A 60/20/20-split was used in each experiments.}
  \label{tbl:removal}
  \begin{tabular}{ccc}
    \hline
    model & \begin{tabular}{@{}c@{}c@{}}RMSE \\ with duplicates \\ (kcal/mol) \end{tabular} & \begin{tabular}{@{}c@{}c@{}}RMSE \\ without duplicates \\ (kcal/mol) \end{tabular} \\
    \hline
    Regular GNN & 10.1 & 10.5 \\
    ml-QM-GNN (full) & 3.9 & 4.0 \\
    ml-QM-GNN (charge + NMR) & 3.9 & 3.9 \\
    ml-QM-GNN (Fukui) & 10.0 & 10.3 \\
    \hline
  \end{tabular}
\end{table}

\section{Note regarding the data curation procedure for the E2/S$_\text{N}$2 classification data set} \label{sec:classification}

While duplicate SMILES did not pose a significant problem for the regression data set (the model accuracies are largely unaffected by their removal or inclusion), the situation for the classification data set is quite different. In its unfiltered version, the classification data set contained 932 data points. These data points correspond to all the reaction systems for which both an E2 and S$_\text{N}$2 transition state was available in the data set, so that a preferential reaction pathway could be assigned. However, several SMILES strings featured in this unfiltered data set are associated with both classes simultaneously (since the two transition states are competitive in energy, so that depending on the TS conformation, one pathway dominates over the other and vice versa). This ambiguity means that irrespective of the specific class assigned to these reacting systems by the model during training, one data point will inevitably have been classified incorrectly. Removing these ambiguous data points altogether (as well as all non-ambiguous duplicates), one retains 791 data points. This is the filtered data set which we have used in our analysis.

\section{Selective sampling results for one-hot encoding in combination with KRR} \label{sec:one-hot encoding}

To assess the performance of one-hot encoding in out-of-sample predictions, the corresponding KRR models developed by Heinen and co-workers were adapted, so that the set of all data points for a single nucleophile is used as the test set, and the data points for all other nucleophiles are used for training (cf. Table \ref{tbl:one-hot-encoding}). \cite{heinen2020quantum} The optimal hyperparameters determined by Heinen et al. were used and no validation set was considered. Note that in the KRR architecture, the competing E2 and S$_\text{N}$2 pathways have to be treated separately (cf. main text). Regardless of these differences, the KRR models generally perform significantly worse than either the regular GNN or ml-QM-GNN models.

\begin{table}
  \caption{Performance -- as expressed in terms of RMSE -- of the one-hot encoding in combination with KRR models in selective sampling for the separated E2 and S$_\text{N}$2 data set. No validation set has been considered; the same hyperparameters as in the work by Von lilienfeld and co-workers have been selected. \cite{heinen2020quantum}}
  \label{tbl:one-hot-encoding}
  \begin{tabular}{ccc}
    \hline
    held-out nucleophile &  RMSE (kcal/mol) for E2 data & RMSE (kcal/mol) for S$_\text{N}$2 data  \\
    \hline
    H$^-$ & 18.62 & 20.18 \\
    F$^-$ & 13.30 & 8.62 \\
    Cl$^-$ & 17.43 & 12.61 \\
    Br$^-$ & 13.94 & 15.48 \\
    \hline
  \end{tabular}
\end{table}

\section{RMSEs and standard deviations obtained for the GNN and ml-QM-GNN during selective sampling} \label{sec:stdev}

\begin{table}
  \caption{RMSEs and standard deviations for the regular GNN in selective sampling of the E2/S$_\text{N}$2 data set. The standard deviations were determined from the five replicates, i.e., the individual iterations of the random train and validation set sampling (cf. main text).}
  \label{tbl:stdev-GNN}
  \begin{tabular}{cc}
    \hline
    held-out nucleophile &  RMSE (kcal/mol)  \\
    \hline
    H$^-$ & $16.32 \pm 1.54$ \\
    F$^-$ & $8.95 \pm 0.29$ \\
    Cl$^-$ & $16.77 \pm 0.27$ \\
    Br$^-$ & $14.47 \pm 0.21$ \\
    \hline
  \end{tabular}
\end{table}

\begin{table}
  \caption{RMSEs and standard deviations for the ml-QM-GNN in selective sampling of the E2/S$_\text{N}$2 data set. The standard deviations were determined from the five replicates, i.e., the individual iterations of the random train and validation set sampling (cf. main text).}
  \label{tbl:stdev-ml-QM-GNN}
  \begin{tabular}{cc}
    \hline
    held-out nucleophile &  RMSE (kcal/mol)  \\
    \hline
    H$^-$ & $8.72 \pm 1.68$ \\
    F$^-$ & $7.67 \pm 1.78$ \\
    Cl$^-$ & $5.35 \pm 0.16$ \\
    Br$^-$ & $5.35 \pm 0.08$ \\
    \hline
  \end{tabular}
\end{table}

\section{The effect of averaging out the respective descriptors for the full ml-QM-GNN model trained on the E2/S$_\text{N}$2 data set}

\begin{figure}[H]
\centering
\includegraphics[scale=0.36]{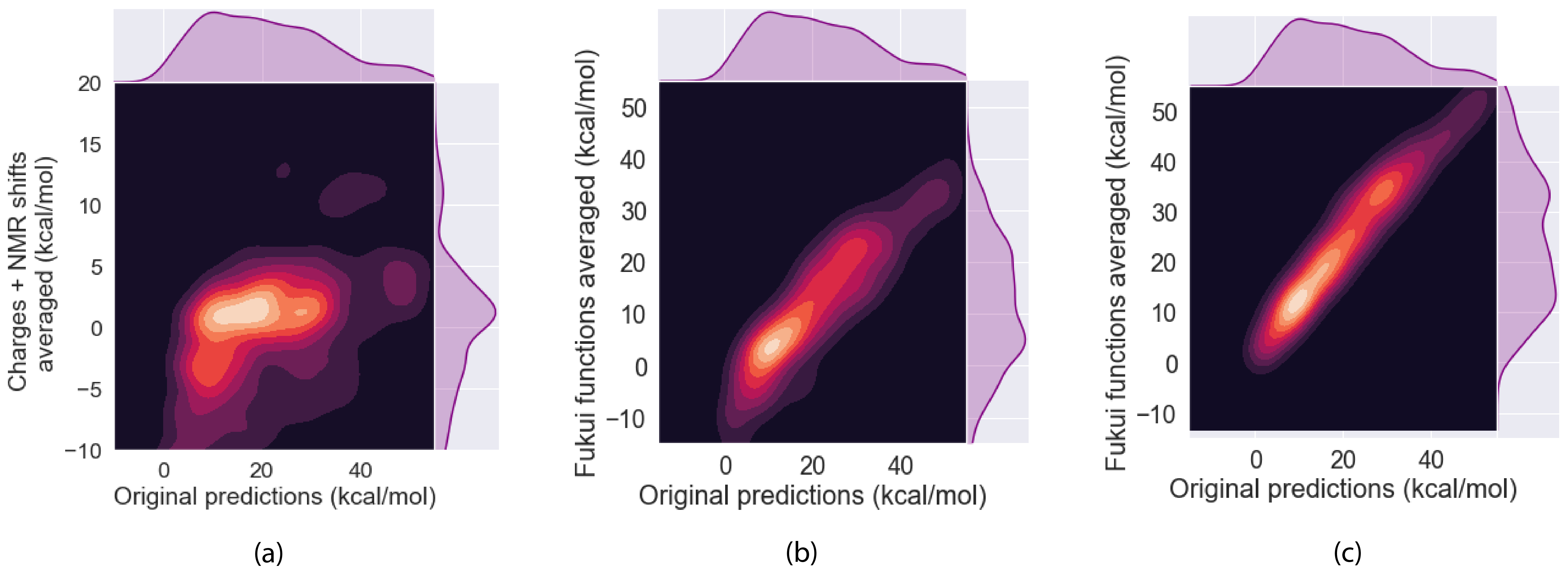}
\caption{(a) Comparison between the original predictions for the full E2/S$_\text{N}$2 activation energy data set made by the trained ml-QM-GNN model with all descriptors included and the predictions made when all charge and NMR shielding constant information is averaged out; (b) comparison between the original predictions for the full data set made by the trained ml-QM-GNN model with all descriptors included and the predictions made when all Fukui function information is averaged out; (c) comparison between the original predictions for the full data set made by the trained ml-QM-GNN model with all descriptors included and the predictions made when all Fukui function information on only the substrates is averaged out (the nucleophiles retain their respective unity Fukui function values).}
\label{fig:scrambled}
\end{figure}

\section{Methodology for the DFT calculations in Figure 6} \label{sec:dft_methodology}

Geometry optimizations of the species were carried out at M06\cite{zhao2008density, zhao2008m06}/def2-TZVP\cite{weigend2005balanced} level-of-theory with the help of the Gaussian09 program.\cite{Pople2003} As explained in the original study, cf. reference \cite{stuyver2021resolving}, the choice of functional was inspired by the results from a benchmarking study by Bento et al.,\cite{bento2008e2} which identified M06 as the hybrid functional leading to reaction profiles agreeing the most with the CCSD(T) reference profiles for competing E2/SN2 reactions (both qualitatively and quantitatively). Corresponding atomic charges were determined from a natural population analysis (NPA).\cite{glendening2013nbo} 

\section{Confusion matrix comparing the classifications by the ml-QM-GNN models to the physical organic criteria for the aromatic substitution data set} \label{sec:confusion}

\begin{landscape}

\begin{table}
  \caption{Aggregated confusion matrix comparing the classifications by the different ml-QM-GNN models with respect to the physical organic criteria for the first 5-fold CV for the nitration reactions and halogenation reactions involving succinimides, present in the aromatic substitution data set. The first column contains all the data points which are correctly classified by the model which considers exclusively hard-hard descriptors, i.e., atomic charges and NMR shielding constants, and are incorrectly classified by the model which considers exclusively soft-soft descriptors, i.e., the electrophilic and nucleophilic Fukui function. The second column contains all the data points which are incorrectly classified by the ml-QM-GNN (charge + NMR) and are correctly classified by the ml-QM-GNN (Fukui). The third column contains those data points for which neither of the two ablated models nor the full ml-QM-GNN model makes the correct prediction, and the fourth column contains all points which are correctly classified by all models. The first and second row in their turn contain those points for which the electrostatic criterion is fulfilled, whereas the (frontier) orbital criterion is not; and vice versa. The third and fourth row respectively contain the points for which neither or all of the criteria are fulfilled. Note that the diagonal elements (highlighted in red) are the ones which are mainly populated, indicating that the predictions made by the respective models tend to adhere to the fulfilment of the corresponding criteria.}
  \label{tbl:testset_concatenated}
  \begin{tabular}{c|cccc|c}
    & \multicolumn{4}{c}{\begin{tabular}{@{}c@{}}comparisons between (ablated) \\ ml-QM-GNN models \end{tabular}} & \\
    \cline{2-5}
    criteria & \begin{tabular}{@{}c@{}}only hard-hard: OK \\ only soft-soft: NOK \end{tabular} & 
    \begin{tabular}{@{}c@{}}only hard-hard: NOK \\ only soft-soft: OK \end{tabular} & all models: NOK & all models: OK & \\
    \cline{1-5}
    \begin{tabular}{@{}c@{}}electrostatic: OK \\ (frontier) orbital: NOK \end{tabular} & \red{53} & 9 & 24 & 107 & \textbf{193}  \\
    \begin{tabular}{@{}c@{}}electrostatic: NOK \\ (frontier) orbital: OK \end{tabular} & 21 & \red{96} & 17 & 308 & \textbf{442} \\
    \begin{tabular}{@{}c@{}}neither criterion \\ fulfilled \end{tabular} & 11 & 11 & \red{52} & 7 & \textbf{81} \\
    \begin{tabular}{@{}c@{}}all criteria \\ fulfilled \end{tabular} & 17 & 48 & 7 & \red{711} & \textbf{783} \\
    \cline{1-6}
    & \textbf{102} & \textbf{164} & \textbf{100} & \textbf{1133} & \textbf{1499} \\
  \end{tabular}
\end{table}

\end{landscape}

\putbib[biblio_ms]
\end{bibunit}

\end{document}